\documentclass[final,5p,twocolumn]{elsarticle}
\usepackage{graphicx}
\usepackage{epsfig}
\usepackage{amssymb}
\usepackage{amsthm}
\usepackage{float}
\usepackage[font={footnotesize,it}]{caption}

\journal{Physics Letters B}

\begin{document}

\begin{frontmatter}

\title{Holographic quantum phase transitions and interacting bulk scalars}

\author[label1]{Pankaj Chaturvedi\corref{cor1}}
\address[label1]{Department of Physics, Indian Institute of Technology Kanpur, Kanpur 208016, India}
\cortext[cor1]{corresponding author}
\ead{cpankaj@iitk.ac.in}

\author[label2]{Pallab Basu}
\address[label2]{International Center for Theoretical Sciences, Tata Institute of Fundamental Research,
Bangalore 560012, INDIA}
\ead{pallab.basu@icts.res.in}

\begin{abstract}
We consider a system of two massive, mutually interacting probe real scalar fields,  in zero temperature holographic backgrounds. The system does not have any continuous symmetry. For a suitable range of the interaction parameters adhering to the interaction potential between the bulk scalars, we have shown that as one turns on the source for one scalar field, the system may go through a second order quantum critical phase transition across which the second scalar field forms a condensate. We have looked at the resulting phase diagram and numerically computed the condensate. We have also investigated our system in two different backgrounds: $AdS_4$ and $AdS$ soliton, and got similar phase structure.
\end{abstract}


\end{frontmatter}

\section{Introduction}
Holography \cite{Maldacena:1997re} or gauge/gravity duality enables us to understand phases of strongly coupled gauge theory from a string theory/gravity calculation. This philosophy has been used in the past decade to model various kind of field theoretic phase transitions in asymptotically anti-de sitter space (AdS). For example it was suggested in \cite{Gubser:2008px}, that the black hole horizons could exhibit spontaneous breaking of an Abelian gauge symmetry if gravity were coupled to matter lagrangian  including a charged scalar (EYMH) that condenses near the horizon. This could be thought of as a superfluid/superconducting like transition in the dual gauge theory \cite{Hartnoll:2008vx}. Such condensed matter inspired system have been studied intensively over the last few years \cite{Herzog:2009xv,Horowitz:2010gk,Sachdev:2010ch}. Similar, EYMH like systems with two or more bulk fields have been studied in \cite{Basu:2010fa,Cai:2013wma,Musso:2013ija,Huang:2011ac,Wen:2013ufa,Nie:2013sda,Amado:2013lia,Amoretti:2013oia,Donos:2013woa,Nishida:2014lta,Li:2014wca}.

To understand the spectrum of possible second order holographic phase transitions, we look at simplistic models which do not even have a local gauge symmetry (conserved particle number in the dual boundary theory). Many condensed matter systems including classic examples like Ising models, have phase transitions which are not necessarily due to breaking of global or local continuous symmetries. In this context the first thing to try is to look at a Lagrangian consisting of a single scalar field. However, such examples are limited. One known example is a scalar field in a near extremal black hole background whose mass is close to BF bound \cite{Hartnoll:2008kx}. Another is by turning on double trace deformations \cite{Faulkner:2010gj}. In this regard we study a generic system with two massive, mutually interacting, real scalar fields  in zero temperature global $AdS_4$ space-time (and also in AdS soliton background) and we establish that such system goes through an interesting quantum critical phase transition. For the system we proceed with turning on the source for one scalar field (which may be thought as an impurity density) and look at the condensation of other field. We find that for suitable attractive interactions between the two bulk scalar fields one may obtain a condensate for the latter field through a second order phase transition.

\section{Model}
In order to describe our model we consider a bulk gravitational action with two mutually interacting and self-gravitating, real scalar fields in asymptotically $AdS_d$ spacetime. The form of bulk gravitational action may be written down as,

\begin{eqnarray} \label{action}
S &=& \frac{1}{2 \kappa^2}\int dx^{d} \sqrt{- g}\left({\cal L}_G \right)+\int dx^{d}\sqrt{- g}\left({\cal L}_M \right)\nonumber\\
{\cal L}_M &=& -\left(\nabla_{\mu}\psi_{1}\right)^2 -\left(\nabla_{\mu}\psi_{2}\right)^2- m_{1}^{2} \psi_{1}^2 - m_{2}^{2}\psi_{2}^2\nonumber\\
&-&  \left(\frac{\alpha_1}{2}\psi_{1}^4+\frac{\alpha_2}{2} \psi_{2}^4\right) + \beta \left(\psi_{1}^2\psi_{2}^2\right),  
\end{eqnarray}
where, $\kappa^2= 8 \pi G$ is related to the gravitational constant in the bulk. This model has a two $Z_2$ symmetries corresponding to $\psi_1 \rightarrow -\psi_1$ and $\psi_2 \rightarrow -\psi_2$.  In principle one may choose a less symmetric model to realize the same phase transition we are looking at. We will be working in the probe limit $(\kappa \rightarrow 0)$ for which we can neglect the back-reaction of the bulk scalar fields on the background geometry. In this limit, by rescaling the equations of the motion for the bulk scalar fields one may argue that phase diagram for the model depends on the ratio of parameters $\frac{\alpha_2}{\alpha_1},\frac{\beta}{\alpha_1}$. We will further assume that, $\alpha_1,\alpha_2 >0$. Also the condition, $\beta < \sqrt{\alpha_1 \alpha_2}$ must be satisfied for the boundedness of the interaction potential. 

Near the boundary the space time look like AdS, and consequently we can expand a scalar field as a leading source ($J$) and a vacuum expectation value term ($< {\cal O} >$). We further assume that $m_1^2=m_2^2<0$. Now when we turn on the source $J_1$ it generates a condensate $\psi_1(x)$. The effective mass of $\psi_2$ is given by,
\begin{equation}
m^{2}_{2,eff}(x)=m_{2}^{2}-\beta \psi_{1}^2(x)
\end{equation}
For  $\beta>0$, as we increase $J_1$ beyond a certain critical value $J_1^c$, the effective mass of $\psi_2$ decreases and a condensation becomes possible for the second bulk scalar field $(\psi_2)$. Alternatively one may say that the formation of a zero mode of $\psi_2=\psi_2^0(x)$ occurs exactly at $J_1=J_1^c$. For the bulk scalar fields it may be noted that as $J_1 \rightarrow \infty$, $\psi_1^2$ approaches the attractor value $-\frac{2 m_1^2}{\alpha_1}$. To have an instability the limiting value of $m^{2}_{2,eff}$ for the scalar field $\psi_2$ must be less than the BF bound. Hence the condition for instability may be written down as,
\begin{equation}
-\beta \frac{2m_1^2}{\alpha_1}> -m^2_{bf}+m_{2}^{2}.
\end{equation}
For $d=4$ and $m_1^2=m_2^2=-2$ we have,
\begin{equation}
\frac{\beta}{\alpha_1}> \frac{1}{16}.\label{boundab}
\end{equation}
Thus to achieve a possible condensation of the second bulk scalar field $(\psi_2)$, the parameters $(\beta,\alpha_1)$ must satisfy the bound (\ref{boundab}) in (3+1) dimensions. 

Secondly, assuming a small condensate $\psi_2=\epsilon \psi^0_2(x)$ and  $J_1=J_1^c+\delta J_1$, it can be estimated that, $\epsilon \sim O(\delta J_1^\frac{1}{2})$. One can also estimate the free energy of the new phase to be negative ($\sim - \epsilon^4$) \cite{Herzog:2010vz,Arean:2010zw}, which is consistent with a second order phase transition. It may also be noted that unlike the holographic superfluid \cite{Hartnoll:2008kx} case, the instability of $\psi_2$ can be dynamical in nature. Thus our model might have implications for studying the quenching dynamics described in \cite{Basu:2011ft}. 

\section{Global AdS Background}
We begin with describing the equations of motion for the bulk scalar fields in the (3+1) dimensional global AdS background. In the probe limit one may write down the global AdS space metric in (3+1) dimensions as,
\begin{equation}
ds^2 = \frac{L^2}{\cos^2x}\left(-dt^2 + dx^2+\sin^2x~d\Omega^2\right),\label{globalmetric}
\end{equation}
where, $d\Omega^2$ is the standard metric on the round unit two-sphere. The ranges of the coordinates are $ − \infty<t<\infty $ and $0\leq x < \pi/2$. We also consider the following ansatz for the two bulk scalars as, 
\begin{equation}
\psi_{1}(x^{\mu})=\psi_{1}(x),~~\psi_{2}(x^{\mu}) = \psi_{2}(x).\label{bulkansatz}
\end{equation}
Now using the metric ansatz (\ref{globalmetric}) and the ansatz for the bulk scalar fields (\ref{bulkansatz}), one may write down the independent equations of motion for the bulk scalar fields $\left(\psi_{1},\psi_{2}\right)$ in the global AdS space background as,
\begin{eqnarray}
\psi _{1}^{''}(x)&=&-\frac{4}{\sin 2x}\psi_{1}^{'}(x)+ m_{1}^2 \sec(x)^2\psi_{1}(x)\nonumber\\
 &+&\sec(x)^2\left(\alpha_1 \psi_{1}(x)^2 - \beta \psi_{2}(x)^2\right)\psi_{1}(x),\label{psi1globalEOM}\\
\psi _{2}^{''}(x)&=&-\frac{4}{\sin 2x}\psi_{2}^{'}(x)+ m_{2}^2 \sec(x)^2\psi_{2}(x)\nonumber\\
&+&\sec(x)^2\left(\alpha_2 \psi_{2}(x)^2 - \beta \psi_{1}(x)^2\right)\psi_{2}(x),\label{psi2globalEOM}
\end{eqnarray} 
here the prime denotes derivative with respect to coordinate $x$. The asymptotic form of the functions $\Theta=\left\lbrace\psi_{1}(x),\psi_{2}(x)\right\rbrace$ near the AdS boundary $x\rightarrow \pi/2$ may be written as,
\begin{eqnarray}
\psi_{1}(x)= J_{1}{\bar x}^{\Delta_{-}}+<{\cal O}_1>{\bar x}^{\Delta_{+}}+\cdots,
\nonumber\\
\psi_{2}(x)= J_{2}{\bar x}^{\Delta_{-}}+<{\cal O}_2>{\bar x}^{\Delta_{+}}+\cdots,
\label{bdryfieldexp}
\end{eqnarray}
where, ${\bar x}=\left(x-\pi/2\right)$ and $\Delta_{\pm}=\left(3\pm\sqrt{9+4 m^{2}}\right)/2$. Here the dots represent the higher order terms in the powers of ${\bar x}$. Now if one takes the masses of scalar fields as $m_1^2=m_2^2=-2$, then from the asymptotic forms of the fields given in equation (\ref{bdryfieldexp}) we observe that one can have dual boundary CFT operators of scaling dimension one $(J_{1},J_{2})$ acting as source terms for the bulk scalar fields. Also we have dual boundary CFT operators of scaling dimension two  $(<{\cal O}_1>,<{\cal O}_2>)$ acting as vacuum expectation values for the bulk scalar fields.

\subsection{Numerical Results}
We attempt to numerically solve the equations of motion via shooting method. Here we require that the functions corresponding to the bulk scalar fields, $\Theta=\left\lbrace\psi_{1}(x),\psi_{2}(x)\right\rbrace$ must be regular at the horizon $(x=0)$. This implies that all the functions must admit finite values and a taylor series expansion near the boundary at $(x=0)$ as,
\begin{equation}
\Theta(x)=\Theta(0)+\Theta'(0) x+\cdots
\end{equation}
Analyzing the equations of motion and the expansions of the functions near $(x=0)$, it may be clearly identified that one must have  two independent parameters at the horizon $(x=0)$, namely $\psi_1(0)$ and $\psi_2(0)$. Out of these two we will use one of them as shooting parameter to get the boundary condition for the sources as $\left\lbrace J_{1}\neq 0, J_{2}= 0\right\rbrace$. The remaining quantities like $<{\cal O}_1>$ and $<{\cal O}_2>$ may be obtained by reading off the corresponding coefficients in the asymptotic forms given in the equation (\ref{bdryfieldexp}) for the bulk scalar fields. The boundary condition on the source terms described above implies that for the numerical analysis we only turn on the source $J_{1}$ for the first scalar field $\psi_1$ and look for the vacuum expectation value $<{\cal O}_2>$ to arise spontaneously for the second scalar field $\psi_2$.

\begin{figure}[H]
\centering
\includegraphics[width=2in,height=1.3in]{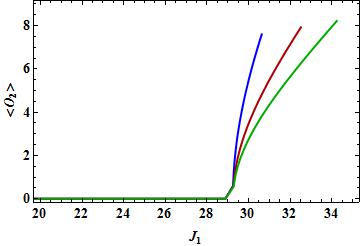}
\caption{\label{fig:GlAdS1}Plots of $<{\cal O}_{2}>$ with respect to $J_1$ for global AdS space-time. The blue, red and the green curves corresponds to different values of the parameter $\alpha_2 =0.4,0.8$ and $1.2$ respectively for fixed values of $\alpha_1 = 1$ and $\beta=0.5$.}
\vspace{0.2cm}
\includegraphics[width=2in,height=1.3in]{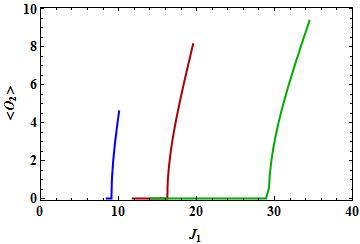}
\caption{\label{fig:GlAdS2}Plots of $<{\cal O}_{2}>$ with respect to $J_1$ for global AdS space-time. The blue, red and the green curves corresponds to different values of the parameter $\alpha_1 =0.4,0.7$ and $1$ respectively for fixed values of $\alpha_2 =1$ and $\beta =0.5$.}
\vspace{0.2cm}
\includegraphics[width=2in,height=1.3in]{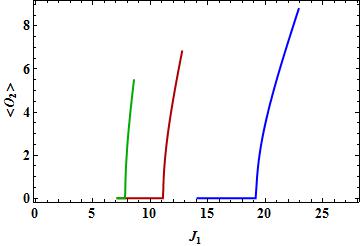}
\caption{\label{fig:GlAdS3}Plots of $<{\cal O}_{2}>$ with respect to $J_1$ for global AdS space-time. The blue, red and the green curves corresponds to different values of the parameter $\beta =0.6,0.8$ and $1$ respectively for fixed values of $\alpha_1 =1$ and $\alpha_2 =1$.}
\end{figure}

For the masses $m_1^2 = m_2^2=-2$ of the bulk scalar fields, in the figures (\ref{fig:GlAdS1},\ref{fig:GlAdS2},\ref{fig:GlAdS3}) we plot the vacuum expectation value $<{\cal O}_{2}>$  with respect to the source $J_{1}$ for varying values of the parameters $\beta,\alpha_1$ and $\alpha_2$. 
From figure (\ref{fig:GlAdS1}) we observe that vacuum expectation value $<{\cal O}_2>$ exists for values of the source $J_1$ above a certain critical value $J_{1c}=28.896$ for varying values of $\alpha_2$. Here the values of the parameters $\alpha_1$ and $\beta$ are kept to be a fixed constant. From figure (\ref{fig:GlAdS2}) we observe that  critical value of $J_1$ increase for increasing values of $\alpha_1$ where the values of the parameters $\alpha_2$ and $\beta$ are kept to be a fixed constant. Similarly  from figure (\ref{fig:GlAdS3}) we observe that  critical value of $J_1$ decreases for increasing values of $\beta$ where the values of the parameters $\alpha_1$ and $\alpha_2$ are kept to be a fixed constant. The most remarkable observation that can be easily interpreted from the figures is that, when we turn on the source $J_1$ for the scalar field $\psi_1$ the second scalar field $\psi_2 $ spontaneously acquires a vacuum expectation value above certain critical value of the source $J_1$. This implies that when one scalar is turned on to act as the source then the interaction between the two bulk scalars forces the other scalar to condense spontaneously which in turn leads to the instability of the global $AdS_4$ bulk.

\section{AdS Soliton Background}
We now consider the two interacting bulk scalar fields in the AdS soliton background. We will still be working in the probe limit which implies that the gravity is non-dynamical. The (3+1) dimensional AdS soliton background may be represented by the following  metric ansatz 

\begin{equation}
ds^2=\frac{dr^2}{f(r)}+r^2\left(-dt^2+dx^2\right)+f(r)d\eta^2, \label{solitonmetric}
\end{equation}
where $f(r)=r^2(1-\frac{r_0^3}{r^3})$ and $r_0$ stands for the tip of the soliton. It may be seen that (\ref{solitonmetric}) is a solution of action described by (\ref{action}) in the absence of any matter source. In order to avoid a conical singularity at $r = r_0$ one must impose the periodicity condition $\eta\rightarrow\eta+\pi/r_0$  on the spatial direction $\eta$. It may be observed that the AdS soliton just looks like a cigar with the asymptotic geometry $R^{1,2}\times S^1$ near the AdS boundary and the spacetime exists only for $r > r_0$. This implies that the boundary field theory dual to the AdS soliton spacetime is in a confined phase with a mass gap. Thus the AdS soliton spacetime represents the insulating phase of the boundary field theory via gauge /gravity duality as pointed out in \cite{Nishioka:2009zj}.  We now consider the following ansatz for the two bulk scalars as, 
\begin{equation}
\psi_{1}(x^{\mu})=\psi_{1}(r),~~\psi_{2}(x^{\mu}) = \psi_{2}(r).\label{bulkansatzSoliton}
\end{equation}

Now for the AdS soliton  background given by the metric in (\ref{solitonmetric}), we may write down the independent equations of motion for the bulk scalar fields $\left(\psi_{1},\psi_{2}\right)$ as
\begin{eqnarray}
\psi_1''(r)&=&-\left(\frac{f'(r)}{f(r)}+\frac{2}{r}\right)\psi_1'(r)+\frac{m_1^2}{f(r)}\psi_1(r)\nonumber\\
&+&\left(\alpha_1\frac{\psi_{1}(x)^2}{f(r)}-\beta\frac{\psi_{2}(x)^2}{f(r)}\right)\psi_1(r),\label{psi1SolitonEOM}\\
\psi_2''(r)&=&-\left(\frac{f'(r)}{f(r)}+\frac{2}{r}\right)\psi_2'(r)+\frac{m_2^2}{f(r)}\psi_2(r)\nonumber\\
&+&\left(\alpha_2\frac{\psi_{2}(x)^2}{f(r)}-\beta\frac{\psi_{1}(x)^2}{f(r)}\right)\psi_2(r),\label{psi2SolitonEOM}
\end{eqnarray} 
here the prime represent the derivative with respect to the coordinate $r$. One may also observe that the metric (\ref{solitonmetric}) and the equations of motion (\ref{psi1SolitonEOM},\ref{psi2SolitonEOM}) posses the following scaling symmetry,

\begin{eqnarray}
r &\rightarrow & \lambda r,~~(t,x,\eta)\rightarrow \frac{(t,x,\eta)}{\lambda} ,\nonumber\\
f &\rightarrow & \lambda^2 f,~~(\psi_1,\psi_2) \rightarrow \lambda (\psi_1,\psi_2),\label{scalingsym}
\end{eqnarray}

\begin{figure}[H]
\centering
\includegraphics[width=2in,height=1.3in]{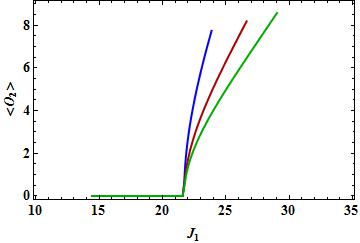}
\caption{\label{fig:SAdS1}Plots of $<{\cal O}_{2}>$ with respect to $J_1$ for AdS soliton background. The blue, red and the green curves corresponds to different values of the parameter $\alpha_2 =0.4,0.8$ and $1.2$ respectively for fixed values of $\alpha_1 = 1$ and $\beta=0.5$.}
\vspace{0.2cm}
\includegraphics[width=2in,height=1.3in]{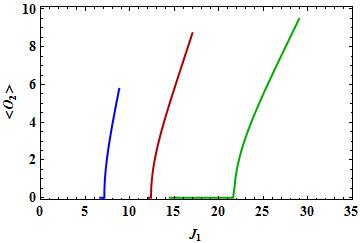}
\caption{\label{fig:SAdS2}Plots of $<{\cal O}_{2}>$ with respect to $J_1$ for AdS soliton background. The blue, red and the green curves corresponds to different values of the parameter $\alpha_1 =0.4,0.7$ and $1$ respectively for fixed values of $\alpha_2 =1$ and $\beta =0.5$.}
\vspace{0.2cm}
\includegraphics[width=2in,height=1.3in]{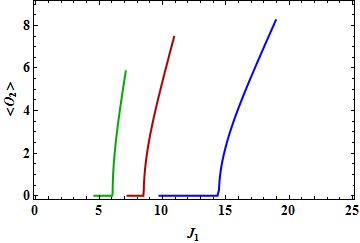}
\caption{\label{fig:SAdS3}Plots of $<{\cal O}_{2}>$ with respect to $J_1$ for AdS soliton background. The blue, red and the green curves corresponds to different values of the parameter $\beta =0.6,0.8$ and $1$ respectively for fixed values of $\alpha_1 =1$ and $\alpha_2 =1$.}
\end{figure}
The scaling symmetry (\ref{scalingsym}) helps to set the parameter $r_0=1$  for performing the numeric computations. Furthermore the asymptotic form of the functions $\Theta=\left\lbrace\psi_{1}(r),\psi_{2}(r)\right\rbrace$ near the boundary at $r\rightarrow \infty$ may be written down as,
\begin{eqnarray}
\psi_{1}(x)&=&J_{1}r^{\Delta_{-}}+<{\cal O}_{1}>r^{\Delta_{+}}+\cdots,\nonumber\\
\psi_{2}(x)&=&J_{2}r^{\Delta_{-}}+<{\cal O}_{2}>r^{\Delta_{+}}+\cdots.\label{bdryfieldexpSAdS}
\end{eqnarray}
where, $\Delta_{\pm}=\frac{3\pm\sqrt{9+4 m^{2}}}{2}$ and the dots represent the higher order terms in the powers of $r$. Here $(J_{1},J_{2})$ act as the source terms for the bulk scalar fields with the dual boundary CFT operators $(<{\cal O}_1>,<{\cal O}_2>)$ acting as the vacuum expectation values for the bulk scalar fields.

\subsection{Numerical Results}
We once again attempt to numerically solve the equations of motion for the case of AdS soliton background via shooting method. In order to begin with the numerics we require that the functions corresponding to the bulk fields, $\Theta=\left\lbrace\psi_{1}(r),\psi_{2}(r)\right\rbrace$ must be regular at the horizon  $(r=r_0=1)$  for the AdS soliton background. This again implies that all the functions must admit finite values and a taylor series expansion near the point $(r=r_0=1)$ as,
\begin{equation}
\Theta(r)=\Theta(1)+\Theta'(1)(r-1)+\cdots,
\end{equation}
Now on analyzing the expansions of the functions near $(r=r_0=1)$, it may be seen that there are two independent parameters at the boundary at $(r=r_0=1)$, namely $\psi_1(1)$ and $\psi_2(1)$. Once again out of these two we will use one as the shooting parameter in order to impose the condition $\left\lbrace J_{1}\neq 0, J_{2} = 0\right\rbrace$ on the asymptotic forms of bulk scalar fields $\psi_1(x)$ and $\psi_2(x)$ near the AdS boundary at $(r\rightarrow \infty)$. The remaining quantities like $<{\cal O}_1>$ and $<{\cal O}_2>$ may be obtained by reading off the corresponding coefficients in the asymptotic forms given in the equation (\ref{bdryfieldexpSAdS}) for the bulk scalar fields. Here also for the numerical analysis we only turn on the source $J_{1}$ for the first scalar field $\psi_1$ and look for the vacuum expectation value $<{\cal O}_2>$ to arise spontaneously for the second scalar field $\psi_2$. 

For the masses $m_1^2 = m_2^2=-2$ of the bulk scalar fields, in the figures (\ref{fig:SAdS1},\ref{fig:SAdS2},\ref{fig:SAdS3}) we plot the vacuum expectation value $<{\cal O}_{2}>$  with respect to the source $J_{1}$ for varying values of the parameters $\beta,\alpha_1$ and $\alpha_2$. In the figures we have chosen the values of the parameters $\alpha_1,\alpha_2$ and $\beta$ to be the same as the values taken for the case of global AdS space-time background. From the figures (\ref{fig:SAdS1},\ref{fig:SAdS2},\ref{fig:SAdS3}) one may observe that the overall behavior of the plots is same as obtained for the case of the global AdS background. The only difference lies in the fact that the critical values of the source $J_1$ for the AdS soliton background are smaller than those obtained for the global AdS background. Thus it seems that one may expect to obtain the signature second order quantum critical phase transition with two mutually interacting bulk scalar fields in any zero temperature AdS background. 

\section{Summary and discussions}
Here we have consider a system of two massive, mutually interacting probe real scalar fields,  in zero temperature holographic backgrounds in order to model a holographic second order quantum phase transition. It seems that in our case when one scalar field condenses, it forces the other scalar field to condense under the mutual interaction. This results in the coexisting condensed phase of the boundary field theory with the nontrivial profiles for both the scalar fields in the bulk global $AdS_4$ space. Thus one may suggest that the system undergoes through an interesting quantum critical phase transition due to the mutual interaction between the bulk scalars. For the two scalar fields we have numerically computed the dual condensates for the boundary operators in the probe limit described above and observed that the system undergoes a second order quantum phase transition. It would also be very interesting to study our system for different masses and interactions (including gravitational) between the two scalar fields, which may lead to interesting novel phase structure of the strong coupled large-$N$ field theory. Another interesting question is to study the system at finite temperature. One may also study time dependent dynamical properties of the system described here. We leave these questions for a future investigation.

\section*{Acknowledgments}
The authors would like to thank Sayantani Bhattacharya and Gautam Sengupta for useful discussions. This work of Pankaj Chaturvedi is supported by the Grant No. 09/092(0846)/2012-EMR-I, from the Council of Scientific and Industrial Research (CSIR), India.

\appendix

\section{A Simple Lagrangian  Model:}
Let us consider a simple Lagrangian model with two mutually interacting scalar fields in $0$ dimension:
\begin{eqnarray}
V(\phi_1,\phi_2) = &\frac{1}{2} (m_1^2\phi_1^2+m_2^2 \phi_2^2+\alpha_1 \phi_1^4+\alpha_2 \phi^4 \nonumber \\
& + \beta \phi_1^2\phi_2^2)+J_1 \phi_1+J_2\phi_2
\end{eqnarray}
The cubic EOMs of this system can be solved in a straightforward manner, but the analytic expressions are too complicated. For the global stability of the potential we have $ \alpha_1 \alpha_2 > \beta^2$.  We broke $\phi_1 \rightarrow -\phi_1$ symmetry by turning on $J_1$ and putting $J_2=0$, so that $\phi_1$ gets a value $\phi_{1s}$. If we expand around $\phi_{1s}$ value the $\phi_2$ mode get a an effective mass $m^2_{eff}(\phi_2)=m_2^2- \beta \phi^2_{1s}$. Hence if $\beta < 0$, then turning on enough $J_1$ would eventually lead a condensation of $\phi_2$. Here a phase transition is always possible irrespective of the value of $\alpha_1$.

Turning on both $J_1$ and $J_2$, by taking linear combination variable may be mapped to a problem where only a single source is turned on. To be kept in mind is that this also changes the potential. For example with  $\alpha_1=\alpha_2=0$ and $\beta>0$ and $m_1=m_2$, a phase transition happens as we gradually turn on $J_1=J_2=J$. The $Z_2$ symmetry between $\phi_1 \leftrightarrow \phi_2$ would be spontaneously broken due to repulsive interaction between $\phi_1$ and $\phi_2$. By using variables $\phi_{\pm}=\frac{1}{2}(\phi_1 \pm \phi_2)$, the potential may be mapped to our standard form $\beta(\phi_+^2-\phi_-^2)^2$ and $J_+=J,J_-=0$.

\bibliographystyle{model1-num-names}

\bibliography{mybib}

\end{document}